\documentclass[12pt,preprint]{aastex}
\shorttitle{GRB050502A Optical Lightcurve}
\shortauthors{Yost et al.}

\begin{document}

\title{Optical Lightcurve \& Cooling Break of GRB~050502A}

\author{
Yost,~S.~A.\altaffilmark{1},
Alatalo,~K.\altaffilmark{1,2},
Rykoff,~E.~S.\altaffilmark{1}, 
Aharonian,~F.\altaffilmark{3},
Akerlof,~C.~W.\altaffilmark{1},
Ashley,~M.~C.~B.\altaffilmark{4}, 
Blake,~C.~H.\altaffilmark{5},
Bloom,~J.~S.\altaffilmark{2},
Boettcher,~M.\altaffilmark{6},
Falco,~E.~E.\altaffilmark{5},
G\"o\u{g}\"u\c{s},~E.\altaffilmark{7},
G\"{u}ver, T.\altaffilmark{8},
Halpern,~J.~P.\altaffilmark{9},
Horns,~D.\altaffilmark{3},
Joshi,~M.\altaffilmark{6},
K{\i}z{\i}lo\v{g}lu,~\"{U}.\altaffilmark{10},
McKay,~T.~A.\altaffilmark{1},
Mirabal,~N.\altaffilmark{1},
\"{O}zel,~M.\altaffilmark{11},
Phillips,~A.\altaffilmark{4}, 
Quimby,~R.~M.\altaffilmark{12},
Rujopakarn,~W.\altaffilmark{1},
Schaefer,~B.~E.\altaffilmark{13}, 
Shields,~J.~C.\altaffilmark{6},
Skrutskie,~M.\altaffilmark{14},
Smith,~D.~A.\altaffilmark{1},
Starr,~D.~L.\altaffilmark{15},
Swan,~H.~F.\altaffilmark{1},
Szentgyorgi,~A.\altaffilmark{5},
Vestrand,~W.~T.\altaffilmark{16}, 
Wheeler,~J.~C.\altaffilmark{12},
Wren,~J.\altaffilmark{16}
}

\altaffiltext{1}{University of Michigan, 2477 Randall Laboratory, 450
        Church St., Ann Arbor, MI, 48104, sayost@umich.edu, kalatalo@umich.edu,
        erykoff@umich.edu, akerlof@umich.edu,
        tamckay@umich.edu, wiphu@umich.edu, donaldas@umich.edu, hswan@umich.edu}
\altaffiltext{2}{UC Berkeley Astronomy, 601 Campbell Hall, Berkeley,
CA, 94720, kalatalo@berkeley.edu, jbloom@astron.berkeley.edu}
\altaffiltext{3}{Max-Planck-Institut f\"{u}r Kernphysik, Saupfercheckweg 1,
        69117 Heidelberg, Germany, Felix.Aharonian@mpi-hd.mpg.de,
        horns@mpi-hd.mpg.de}
\altaffiltext{4}{School of Physics, Department of Astrophysics and Optics,
        University of New South Wales, Sydney, NSW 2052, Australia,
        mcba@phys.unsw.edu.au, a.phillips@unsw.edu.au}
\altaffiltext{5}{Harvard College Observatory, Cambridge, Massachusetts 02138, cblake@cfa.harvard.edu, efalco@cfa.harvard.edu, aszentgyorgyi@cfa.harvard.edu}
\altaffiltext{6}{Ohio University, Athens, OH 45701, boettchm@ohio.edu, Manasvita.Joshi.1@ohio.edu, shields@phy.ohiou.edu}
\altaffiltext{7}{Sabanc{\i} University, Orhanl{\i}-Tuzla 34956 Istanbul, Turkey, ersing@sabanciuniv.edu}
\altaffiltext{8}{Istanbul University Science Faculty, Department of Astronomy
        and Space Sciences, 34119, University-Istanbul, Turkey, 
        tolga@istanbul.edu.tr}
\altaffiltext{9}{Columbia University, Columbia Astrophysics Lab, 550 W. 120th St. Mail Code 5230, New York, NY 10027-6601, jules@astro.columbia.edu}
\altaffiltext{10}{Middle East Technical University, 06531 Ankara, Turkey,
        umk@astroa.physics.metu.edu.tr}
\altaffiltext{11}{\c{C}anakkale Onsekiz Mart \"{U}niversitesi, Terzio\v{g}lu
        17020, \c{C}anakkale, Turkey, me\_ozel@ibu.edu.tr}
\altaffiltext{12}{Department of Astronomy, University of Texas, Austin, TX
        78712, quimby@astro.as.utexas.edu, wheel@astro.as.utexas.edu}
\altaffiltext{13}{Department of Physics and Astronomy, Louisiana State
        University, Baton Rouge, LA 70803, schaefer@lsu.edu}
\altaffiltext{14}{University of Virginia, Department of Astronomy,
   PO Box 3818,
   Charlottesville, VA 22903, mfs4n@virginia.edu}
\altaffiltext{15}{Gemini Observatory, Hilo, Hawaii 96720, USA, dan@pairitel.org}
\altaffiltext{16}{Los Alamos National Laboratory, NIS-2 MS D436, Los Alamos, NM
        87545, vestrand@lanl.gov, jwren@nis.lanl.gov}

\begin{abstract}

We present lightcurves of the afterglow of GRB~050502A, including very
early data at $t-t_{GRB} < 60$~s.  The lightcurve is composed of
unfiltered ROTSE-IIIb optical observations from 44~s to 6~h
post-burst, $R$-band MDM observations from 1.6 to 8.4~h post-burst,
and PAIRITEL $J H K_s$ observations from 0.6 to 2.6~h post-burst.  The
optical lightcurve is fit by a broken power law, where $t^{\alpha}$
steepens from $\alpha = -1.13 \pm 0.02$ to $\alpha = -1.44 \pm 0.02$
at $\sim$5700~s. This steepening is consistent with the evolution
expected for the passage of the cooling frequency $\nu_c$ through the
optical band.  Even in our earliest observation at 44~s post-burst,
there is no evidence that the optical flux is brighter than a backward
extrapolation of the later power law would suggest. The observed decay
indices and spectral index are consistent with either an ISM or a Wind
fireball model, but slightly favor the ISM interpretation. The
expected spectral index in the ISM interpretation is consistent within
$1\sigma$ with the observed spectral index $\beta = -0.8 \pm 0.1$; the
Wind interpretation would imply a slightly ($\sim2\sigma$) shallower
spectral index than observed.  A small amount of dust extinction at
the source redshift could steepen an intrinsic spectrum sufficiently
to account for the observed value of $\beta$.  In this picture, the
early optical decay, with the peak at or below $4.7\times10^{14}$~Hz
at 44~s, requires very small electron and magnetic energy partitions
from the fireball.

\end{abstract}

\section{Introduction}

GRB afterglows are typically observed to decay as power laws in time
\citep[as reviewed by, e.g.,][]{piran04}. The leading afterglow model
is the synchrotron fireball \citep{mr97,spn98}. It describes the
afterglow as synchrotron emission from shock-accelerated electrons
with a broken power law spectral energy distribution, with several
characteristic break frequencies. When the typical synchrotron
frequency ($\nu_m$) or the cooling frequency ($\nu_c$) passes through
the optical bands, the model predicts a break in the lightcurve. 

The fireball model's spectral breaks have different power law indices
depending upon the circumburst medium \citep[as discussed
by][]{mrw98}. The different resulting lightcurves allow important
physical distinctions to be inferred from early afterglow
observations. One example is the anticipated difference between a
constant circumburst density (called ``ISM'') or $r^{-2}$ density
gradient (called ``Wind'' as it resembles the environment produced by
a steady mass-loss wind outflow). The GRB follow-up community has
worked to produce ever-earlier observations in order to detect
lightcurve breaks.

The passage of the cooling frequency, $\nu_c$, has been inferred from
changes in the broadband spectral index. The first such example was
GRB~970508 \citep{gwbgs+98}, where the optical-to-X-ray slope
steepened as if $\nu_c$ were passing between them. Direct evidence of
a cooling break passage in lightcurves is rarer. GRB~030329 showed a
shallow change of slope at $t < 0.25$~d postulated to be the cooling
break \citep{sksyk+03}, although the standard fireball picture is
complicated in that case, possibly with ``layered'' ejecta giving 2
jets and 2 jet breaks \citep{bkpfm+03}. More recently,
\citet{hufhi+05} found a shallow break in the GRB~040924 afterglow
better explained by $\nu_c$ than a jet break.

Here we report observations of the GRB~050502A afterglow in unfiltered
optical and $R J H K_s$ bands. The optical data span nearly 3
logarithmic decades in time and begin at $t < 1$~min after the start
of the GRB. We discuss the lightcurves in the context of the fireball
model.

\section{Observations}\label{sec:observations}

This paper's optical and near-IR lightcurves are the result of three
observation teams with different instruments at separate sites:
ROTSE-III, PAIRITEL, and MDM.

The ROTSE-III array is a worldwide network of 0.45~m robotic,
automated telescopes, built for fast ($\sim 6$ s) responses to GRB
triggers from satellites such as HETE-2 and {\em Swift}.  They have a
wide ($1\fdg85 \times 1\fdg85$) field of view imaged onto a Marconi
$2048\times2048$ back-illuminated thinned CCD, and operate without
filters, with a bandpass from approximately 400 to 900 nm. ROTSE-IIIb
is located at McDonald Observatory in Texas. The ROTSE-III systems are
described in detail in \citet{akmrs03}.

The 1.3m diameter PAIRITEL (Peters Automated Infrared Imaging
Telescope) is a fully automated incarnation of the system used for the
Two Micron All-Sky Survey (2MASS). It is located on the Ridge at
Mt. Hopkins, Arizona. The camera consists of three 256$\times$256
NICMOS3 arrays that image simultaneously the same $8.\!^{\prime}5
\times8.\!^{\prime}5$ portion of the sky at $J$, $H$, and $K_s$ bands
(central wavelengths 1.2, 1.6, and 2.2 $\mu$m).

The MDM Observatory is located at Kitt Peak, Arizona. It includes the
1.3m McGraw-Hill telescope, which covers an $8.\!^{\prime}7 \times
8.\!^{\prime}7$ field of view. This instrument operates with a
standard set of filters. The camera is a SITe $1024 \times 1024$
thinned, backside illuminated CCD, with a pixel scale of
$0.\!^{\prime\prime}508$.

On 2005 May 2, INTEGRAL detected GRB~050502A (INTEGRAL trigger 2484) at
02:13:57 UT. The position was distributed as a Gamma-ray burst
Coordinates Network (GCN) notice at 02:14:36 UT, with a $3\arcmin$ radius error
box, 39~s after the start of the burst \citep{gmmsb05}. The burst had a
duration of 21~s, with a fluence of
$1.4\times10^{-6}\,\mathrm{erg}\,\mathrm{cm}^{-2}$ in the 20-200 keV
band \citep{gm05}.

ROTSE-IIIb responded automatically to the GCN notice in 5.0~s with the
first exposure starting at 02:14:41.0 UT, 44~s after the burst and
only 23~s after the cessation of $\gamma$-ray activity.  The automated
scheduler began a program of ten 5-s exposures, ten 20-s exposures,
and 412 60-s exposures. The first exposure was taken during evening
twilight hours, and ROTSE-IIIb was able to follow the burst position
until the morning twilight. The early images were affected by
scattered clouds and a bright sky background.  Near real-time analysis
of the ROTSE-III images detected a $14^{th}$ magnitude fading source
at $\alpha=13^h29^m46\fs3$, $\delta=+42\arcdeg40\arcmin27\farcs7$
(J2000.0) that was not visible on the Digitized Sky
Survey\footnote{http://archive.stsci.edu/cgi-bin/dss\_form} red
plates, which we reported via the GCN Circular e-mail exploder within
40 minutes of the burst~\citep{yssa05}.

The PAIRITEL instrument on Mt. Hopkins received the GRB~050502A
trigger before dusk and generated a rapid response ToO. When an
interrupt is generated by a new burst at nighttime, a new set of
observations is queued, overriding all other targets. Typical time
from GCN alert to slew is 10 seconds and typical slew times are 1-2
minutes. Since this burst occurred before nighttime science
operations, it overrode the scheduled observations related to
telescope pointing. This behavior yielded a bad pointing model in the
first set of observations and has subsequently been prohibited. The
first set of usable imaging began at 02:52:39.5 UT, or approximately
39 minutes after the trigger. Several imaging epochs were conducted
over the following 5 hours.

PAIRITEL images were acquired with double-correlated sampling with effective
exposure times of 7.848 seconds and dithered over several positions
during a single epoch of observation. Typical total integration times
are between 2 to 30 minutes with approximately 10--30 different dither
positions. 

The MDM Observatory began $R$-band observations 1.8 hours after the
burst, following the initial ROTSE GCN report. 21 exposures were taken
of the GRB field spanning a total of 6.4 hours.

\section{Data Reductions}\label{sec:reductions}

The three diverse datasets required different photometric
reductions. All the data described below are given in Table
\ref{tab:photometry}.

The ROTSE-IIIb images were dark-subtracted and flat-fielded with its
standard pipeline.  The flat-field was generated from 30 twilight
images. SExtractor~\citep{ba96} was applied for the initial object
detection.  The images were then processed with a customized version
of the DAOPHOT PSF fitting package~\citep{stetson87} that has been
ported to the IDL Astronomy User's Library~\citep{landsman95}.  The
PSF is calculated for each image from a set of $\sim50$ well-measured
stars within $12'$ of the target location.  The PSF is fit
simultaneously to groups of stars using the {\tt nstar}
procedure~\citep{stetson87}. Relative photometry was then performed
using 13 neighboring stars. 

ROTSE-III magnitudes are calibrated as $R$-equivalent to accomodate
ROTSE's peak optical bandpass sensitivity at red wavelengths. The
$R$-equivalent magnitude zero-point is calculated from the median
offset to the USNO~1~m $R$-band standard stars~\citep{henden05} in the
magnitude range of $13.5<V<20.0$ with the color of typical stars in
the field such that $0.4 < V-R < 1.0$.  As we have no data on
afterglow color information at the early time, no additional color
corrections have been applied to the unfiltered data, and the
magnitudes quoted are then treated as $R$-band and referred to as
``$C_{R}$''.

The PAIRITEL data were processed through its custom reductions
pipeline and then combined into mosaics as a function of epoch
and filter. The mosaicking used a cross-correlation between
reduced images to find the pixel offsets for the dithers. Final images
were constructed using a static bad pixel mask and drizzling
\citep{fh97}. Bias and sky frames are obtained for each frame by
median combining several images before and after that frame. Flat
fields are applied from archival sky flats, and are known to be highly
stable over long periods of time.  Aperture photometry was performed
in a 3.5 pixel radius aperture (the average seeing FWHM was 2.4
pixels). The PAIRITEL photometric zeropoints for each of $J$, $H$, and
$K_s$ were determined from 2MASS catalog stars, of which there are
$\sim 20$ in the field.

The MDM data were processed using standard IRAF/DAOPHOT procedures.
Aperture photometry was performed in a $1.5\!^{\prime\prime}$ radius
aperture (average seeing was $\sim$ $1.5\!^{\prime\prime}$) centered
on the OT and nearby field stars. Instrumental magnitudes were then
transformed to the $R$ system using the latest calibration provided by
Henden et al. (2005), using differential photometry for each image
with respect to two Henden stars, the only ones available in the MDM
field (RA, DEC = 13:30:04.3, +42:41:06.0 and 13:29:59.8,+42:43:00.2,
both J2000).

\section{Results}\label{results}

In order to analyze the results, we convert all magnitudes from Table
\ref{tab:photometry} to spectral flux densities. For the $R$ (or ROTSE
$C_R$) magnitudes, we use the effective frequencies and zeropoint
fluxes of \citet{b79}. The $JHK_s$ magnitudes are converted using
\citet{cwm03}. There is scant Galactic
extinction at the high Galactic latitude of the transient. However, we
do correct the fluxes for 0.028, 0.01, 0.006, and 0.004 mags of
extinction in the $R$ (\& $C_R$), $J$, $H$, and $K_s$ bands
respectively, as found from the extinction estimates of \citet{sfd98}.

\citet{pefbc05} established the source redshift, $z=3.793$. At such
high $z$, absorption from the Lyman-$\alpha$ forest becomes important
in the ROTSE-III bandpass. The bandpass covers the wavelength range of
approximately $B$ to $I$, but the Ly$\alpha$ absorption from this $z$
significantly depresses wavelengths corresponding to $B$ and
$V$. Ordinarily the blue spectrum of an OT would cause an
overestimation of the $C_R$-band flux from ROTSE instrumental magnitudes,
but in this case extinction across a significant fraction of the
bandpass makes the conversion using the zeropoint from unextincted
field stars {\em underestimate} the $C_R$-band flux. As a result, we
expect a color term between the ROTSE $C_R$ and the MDM $R$
magnitudes. The precise color term is a complicated function of the
Ly$\alpha$ absorption and the OT spectrum.  We fit a constant color
term parameter when examining $R$ and $C_R$ data together, as a
constant multiplicative offset of the ROTSE points relative to the MDM
data. This constancy presumes that any OT color changes within
the ROTSE bandpass have a small effect upon the color term parameter
by comparison to the data uncertainties.

\begin{figure}
\plotone{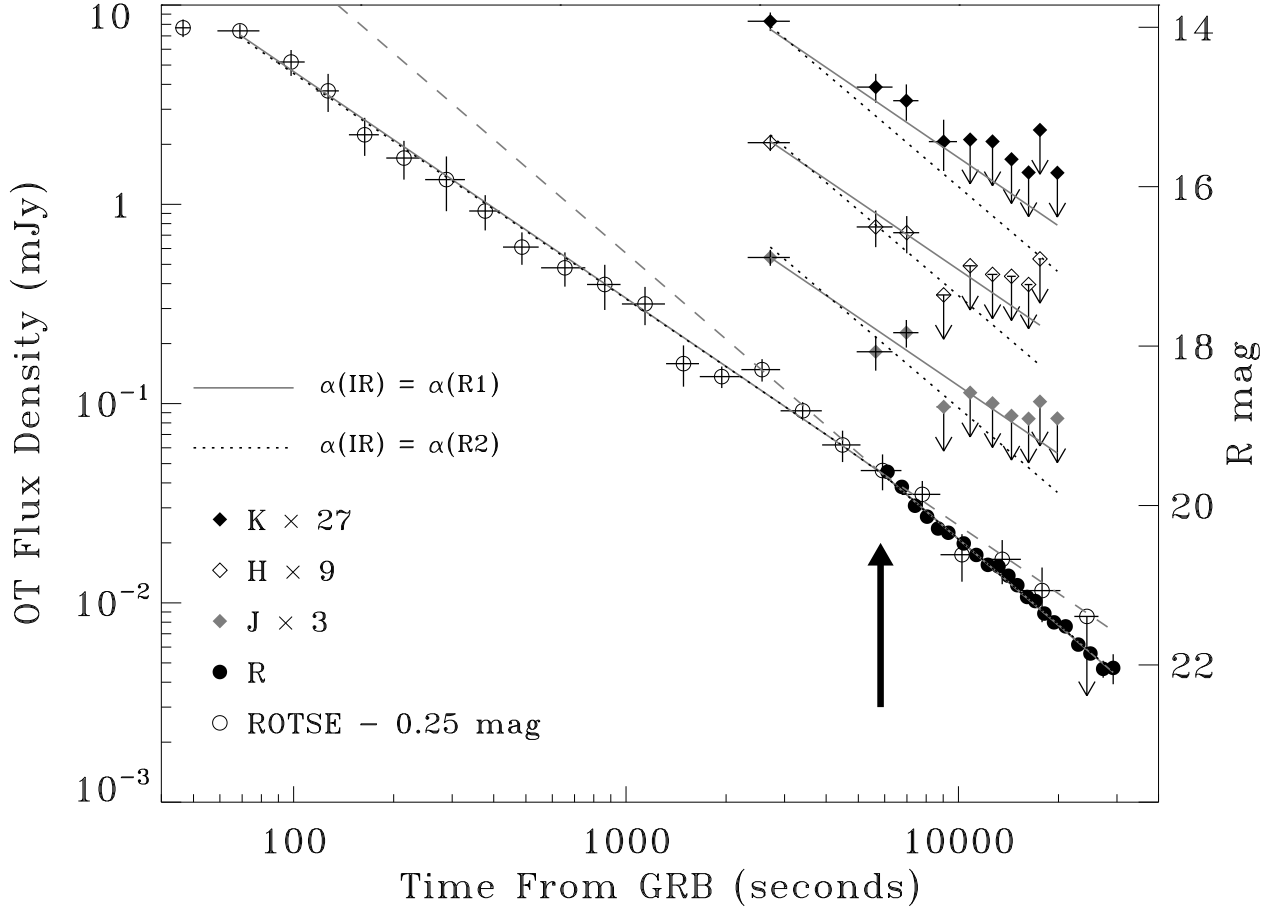}
\caption{ Lightcurves of the optical-IR data presented in this
paper. These are converted to fluxes as described in \S\ref{results},
and the $J$, $H$, and $K_s$ bands are offset by factors of 3, 9, and
27 respectively for clarity. The ROTSE points are unfiltered and
calibrated to $R$ ($C_R$). These points are offset by a color term of
0.25 magnitudes, a value determined from the $R$/$C_R$-band fit (see
\S\ref{results}). Two fit models from Table~\ref{tab:fitpars} are
overplotted. Each fits the bands simultaneously, scaling flux density
$f_{\nu}\propto \nu^{\beta} t^{\alpha}$, with a break in $R$. The
magnitude of the $R$ lightcurve break matches the passage of the
cooling frequency $\nu_c$. The solid fit requires $\alpha(IR) =
\alpha_{R1}$, the initial $R$-band decay, and the dotted fit requires
$\alpha(IR) = \alpha_{R2}$, the final $R$-band decay. The dashed grey
lines show the extrapolations of the initial and final $R$ temporal
power laws (solid fit), and the arrow denotes the break time. Both
fits are reasonable; $\alpha(IR) = \alpha_{R1}$ is expected for an ISM
model, and $\alpha(IR) = \alpha_{R2}$ for a Wind. However, the
$\alpha(IR) = \alpha_{R2}$ fit underestimates the $K_s$
data.}\label{fig:fig1}
\end{figure}

The lightcurves are plotted in Figure \ref{fig:fig1}. The $R$-band
lightcurve appears to be a broken power law. The need for a break is
made obvious by the precise MDM observations at later times. The first
(ROTSE) point in the lightcurve at $t - t_{GRB} = 44$~s seems
underluminous relative to subsequent decay. The IR data are less
well-sampled, but can be described as a set of power laws with decay
rates similar to those seen in the $R$ band.

To quantify these trends, we fit the data using power law and broken
power law forms. The smooth broken power law form is
\begin{equation}
f(t) = f_0 2^{1/s} ( (t/t_b)^{-s \alpha_1} + (t/t_b)^{-s \alpha_2} )^{-1/s},
\end{equation}
where $t_b$ is the break time, $f_0$ is the flux at the break time,
$\alpha_1$ and $\alpha_2$ are the two power law indices, and $s$ is a
sharpness parameter. The data (Fig.~\ref{fig:fig1}) clearly require a
very sharp break transition. We use the form above for convenience,
but do not fit the sharpness, setting $s$ at a very high value ($s=125$).
As noted
above, we also fit a constant multiplicative factor for the ROTSE
$C_R$ points, as a color term. The ROTSE points after the
break observed in the data have significant uncertainties. We expect
any color change that may be associated with the break will not change
the color term significantly in comparison to the ROTSE error bars.

The ``separate'' fit uses the form discussed above for the $R$-band
data and individual power laws for each IR band (there is no substantial
evidence for breaks in the IR data). This allows each band to have
different temporal decays and flux normalizations.

The ``multiband'' fit connects the wavelength bands via a spectral
index and coupled temporal decays, as expected in the fireball
model. This fit uses the above functional form for $R$-band data, and
simultaneously fits single power laws for the IR of the form
\begin{equation}
f(\nu_i,t) = f_{0i} (\nu_i/\nu_{R})^{\beta} (t/t_b)^{\alpha_{IR}},
\end{equation}
where $\nu_i$ designates the frequency of an IR band. We attempt two
``multiband'' forms, $\alpha_{IR} = \alpha_{1}$ (the initial $R$-band
decay index) and $\alpha_{IR} = \alpha_{2}$ (the final $R$-band decay
index).

\subsection{Early Underluminosity?}\label{earlyunderlum}

First, we examined the $R$-band lightcurve (ROTSE \& MDM). A single
power law does not produce an acceptable fit.  We fit all $R$-band
data (including ``$C_R$'') to the ``broken power law + color term''
function described above, and repeated the exercise excluding the
first point. Both produced formally acceptable fits, but excluding the
first point improved the fit notably from $\chi^2 =52.8$ for 39
degrees of freedom (DOF) to $\chi^2 = 44.9$ for 38 DOF.

The first observation, at 44~s post-GRB, appears to plateau, not
joining the overall decay seen later. The flux density is
approximately 4$\sigma$ below the back-extrapolation of the fit
excluding the first point, marginal evidence for a deviation at early
times from the later decay. While a single point cannot definitively
establish such a deviation, the first point does not fit the overall
later decay and thus we exclude it from the lightcurve fits discussed
below and recorded in Table~\ref{tab:fitpars}. The potential
constraints placed by this first point are discussed later, in
\S\ref{sec:fireball}.

Overluminosity, or flux above the expectations from the subsequent decay,
would be expected if a reverse shock's emission were observable as it
passes through the ejecta at early times comparable to our first
observation \citep[e.g.,][]{mr97,sp99a}. This is certainly not seen in
our lightcurve. The maximum flux we observe in the first observation
(99\% confidence upper limit) is only 90\% of the back-extrapolated
flux from the fit excluding the first data point. A flux level
significantly above this extrapolation would have been readily
measurable.

\subsection{Fitting the Frequency Bands Separately}\label{separatefit}

The $R$ (\& $C_R$) band, excluding the first point, were fit to the
broken power law as described above, and each of the $J$, $H$, and
$K_s$ bands were independently fit to a power law. Nondetections are
included as flux densities of $0 \pm f(m_{lim})$ to force the fit not
to overestimate undetected fluxes. However, the elimination of
nondetections from the fits does not change the parameters
significantly.  The resulting fit parameters are given in Table
\ref{tab:fitpars}. Overall, the total $\chi^2 = 60$ for 61 DOF.

Due to the Ly$\alpha$ absorption discussed above, the $R$-band fit
requires a significant color term, 0.25 magnitudes, for the ROTSE
$C_R$ data relative to the filtered MDM $R$ measurements. With this
systematic correction, the $R$-band lightcurve has a sharp break at a
time $5700\pm800$~s post-burst. The $R$-band lightcurve break is from
a decay index $\alpha$ of $-1.13 \pm 0.02$ to $-1.44 \pm 0.02$, a
$\Delta\alpha = -0.31 \pm 0.03$. In the fireball model, the expected
steepening from a ``jet break'' (observing a lack of flux due to the
edge of conical ejecta) is at least $\Delta\alpha = -0.75$
\citep{pm99,sph99}. The only shallow steepening expected in the
fireball model is from the passage of the cooling frequency. The model
predicts $\Delta\alpha(\nu_c=\nu) = -0.25$, consistent with the
observed break at the $1.8 \sigma$ level.

The NIR data are consistent with power law fits. The fitted $J$, $H$,
and $K_s$ decay rates lie between the initial and final $R$-band decay
rates. They are internally consistent at the 1$\sigma$ level and thus
show no evidence for NIR color changes. The NIR decay indices are not
as tightly constrained as those describing the $R$-band, with
individual uncertainties $\delta\alpha\approx0.2$. As the $R$-band
break is shallow, each NIR band's decay is consistent with either the
initial or the final $R$-band decay index (at the 1.6$\sigma$ level or
better).

Consistency with the {\em initial} $R$-band decay would be expected
for an ISM fireball in which $\nu_c$ drops as $\propto t^{-1/2}$,
while consistency with the {\em final} $R$-band decay would be
expected for a Wind fireball where $\nu_c$ rises as $\propto t^{1/2}$
\citep[as reviewed by][]{piran04}. The $K_s$ band is slightly more
consistent with the ISM picture than the Wind, the $H$ and $J$ with
the Wind than the ISM. 

As evident from the poorly-determined NIR decay rates, the NIR data do
not have the sensitivity to determine the presence or absence of a
lightcurve break as shallow as that observed in the
$R$-band. Moreover, the NIR time coverage is from 2320~s to 20700~s
and the observed $R$-band break is at $5700\pm800$~s. Even if the NIR
data were more sensitive, given the timing of the $R$-band lightcurve
break, in the ISM (Wind) model prediction, $\nu_c$ would not be
observed falling (rising) to (from) the $J$ band during the NIR
coverage. The ISM model would predict that $\nu_c$ would fall from $R$
to $J$ only at $t \sim 21000\pm3000$~s. The Wind model would predict
that $\nu_c$ would rise from $J$ to $R$, with the break in $J$ at $t
\sim 1600\pm200$~s.

\subsection{Multiband Fitting}

Given no obvious color changes, we also fit the multiband function as
described in \S\S\ref{earlyunderlum}, \ref{separatefit}. We gave
initial parameters to search for two cases, with the NIR decay either
the power law behavior of the initial $R$-band lightcurve decay, or
of the final $R$ decay. The two results are shown in Table
\ref{tab:fitpars}. The $R$ decay indices, the $R$-band break time, and
the $R$ flux at the break time do not change
significantly. The break amplitude $\Delta\alpha$ is
consistent with the cooling break passage at the 1.5$\sigma$ level in
both cases.

Both multiband fits are formally quite good, with
$\chi^2$/DOF$\approx$1. The data are insufficiently constraining to
distinguish between $\alpha_{IR} = \alpha_{R1}$ (ISM expectation), and
$\alpha_{IR} = \alpha_{R2}$ (Wind expectation). We show both of these
fits in Figure~\ref{fig:fig1}, where the shallower $\beta$ favored by
the $\alpha_{IR} = \alpha_{R2}$ model and the steeper IR decay make
the fitted model underestimate all the $K_s$ data slightly. As the fit
is nevertheless good, this only mildly favors $\alpha_{IR} =
\alpha_{R1}$ and the ISM interpretation.

\subsection{Spectral Index $\beta$}\label{sec:beta}

The optical-NIR spectrum of the data is well fit by a power law
$f_{\nu} \propto \nu^{\beta}$.  This is observed in the NIR data alone
($JHK_s$ at one common epoch), as well as $RJHK_s$ data (interpolating
the $R$-band point). There is no significant change across observation
epochs as there is no evidence for color changes from the
lightcurves. The temporal decays of the IR bands are not strongly
constrained ($\Delta\alpha\approx0.2$, Table~\ref{tab:fitpars}), are
all mutually consistent, and are consistent (at $< 2\sigma$) with either the
initial or the final $R$-band decay.  Moreover, the complete dataset
is well fit by the ``multiband'' function that includes the spectral
index $\beta$ to connect the bands.

The $JHK_s$ spectrum at the initial epoch yields $\beta=-0.9 \pm 0.3$,
while $RJHK_s$ gives $\beta=-0.9 \pm 0.2$. These are not as
well-constrained as the $\beta$ fit using the ``multiband''
function. When $\alpha_{IR} = \alpha_{R1}$, $\beta=-0.79\pm0.05$ is
fit, and when $\alpha_{IR} = \alpha_{R2}$, $\beta=-0.64\pm0.06$.

These results are all mutually consistent. Given their variation, we
take our measure of the optical-NIR spectral index to be $\beta = -0.8
\pm 0.1$. The observed $R$-band lightcurve break will produce color
changes relative to the single power laws of the NIR bands, but they
are not detectable as we do not have good IR detections in the period
from the break to the end of our NIR observations.

\section{Discussion}\label{sec:discuss}

The power law break in the $R$-band lightcurve is consistent with
$\Delta\alpha=0.25$, the value expected for the passage of a cooling
break $\nu_c$ in a fireball model with either an ISM or Wind
circumburst density.  The fireball model can also steepen a decay via
a jet break, or via the passage of the typical frequency $\nu_m$ in
the fast-cooling ($\nu_c<\nu_m$) case. The latter, for an observing
frequency $\nu$, is the transition from $\nu_c < \nu < \nu_m$ to
$\nu_c < \nu_m < \nu$.  As previously noted, a jet break should be
significantly steeper than the $\Delta\alpha \approx 0.3$ observed
\citep{pm99,sph99}. The $\nu_m$ passage in a fast-cooling case
requires the initial decay to be $t^{-1/4}$ before the break
\citep[\S~VIIB]{piran04} which is not compatible with any of the
lightcurves.

From the temporal indices, we conclude that we are observing the
passage of $\nu_c$ through $R$ at $t \approx 6000$~s. In the ISM case,
this is a passage from $\nu_m < \nu(R) < \nu_c$ to $\nu_m < \nu_c <
\nu(R)$. In the Wind case, it is from $\nu_m < \nu_c < \nu(R)$ to
$\nu_m < \nu(R) < \nu_c$. Both require that $\nu_m$ is below the
optical by the first fit data point at $t-t_{GRB} = 59$~s.

\subsection{Interpreting $\alpha$--$\beta$}\label{sec:alphabeta}

The fireball model predicts relations between observed temporal and
spectral indices and the value $p$ ($N(\gamma)\propto\gamma^{p}$) of
the power law index for the spectral energy distribution of
synchrotron-emitting electrons. The relations are functions of
synchrotron spectral break ordering, the spectral segment, and whether
the fireball is the ISM or Wind case, compiled by
\citet{piran04}. With an initial and a final decay index, there are
two relations to determine $p$ for each assumed case (ISM or
Wind). Using the $R$-band decay indices from the ``separate'' fit
gives values of $p$, $2.51 \pm 0.03$ (from the initial $\alpha_{R1}$)
and $2.58 \pm 0.03$ (final $\alpha_{R2}$), for the constant-density
ISM case, and $2.17 \pm 0.03$ ($\alpha_{R1}$) and $2.25 \pm 0.03$
($\alpha_{R2}$), assuming the Wind case.

Given the relations between $p$ and $\alpha$, as well as those between
$p$ and $\beta$, $\alpha$ and $\beta$ are related \citep[again
compiled in][\S~VII]{piran04}. With the $R$ lightcurve break modelled
by the passage of $\nu_c$ in either an ISM or a Wind-like medium, there
are four possible $\alpha$--$\beta$ ``closure'' relations of
$g(\alpha,\beta)=0$. We summarize them in Table~\ref{tab:alphabeta},
along with the values for the electron energy spectral index
$p(\alpha)$ in these cases, and $p(\beta)$.

As before, the ISM model better represents the data. The ISM closure
$g(\alpha,\beta)$ results are more consistent with 0, but the Wind
cases do not deviate by more than $\approx2\sigma$. Similarly, the values of
$p$ inferred from the ISM relations $p(\alpha)$ better agree with its
$p(\beta)$ than in the Wind case, with a similar level of consistency
as the closure relations.

The observed spectral index of $-0.8 \pm 0.1$ is consistent with
either the ISM or Wind picture, as seen in the closure relations, but
is slightly steeper than the Wind model expects
(Table~\ref{tab:alphabeta}), by $\Delta\beta \approx 0.2$.

At the source redshift of $z=3.793$ \citep{pefbc05}, the $K_s$-band in
the observer frame is roughly $B$-band in the local frame, and the
$R$-band is near-UV at $2\times 10^{15}$~Hz. Over this frequency range
in the source frame, a dust extinction law such as that of the Large
Magellanic Cloud (LMC) will be linear, leading to a simple power law
steepening of the spectrum \citep{fm88, reichart01}.  Dust extinction
such as that observed in the Milky Way would have a significant bump
in the middle of that frequency range, so we did not consider it. We
use the LMC prescription of \citet{reichart01} and determine that a
small local extinction $A(V) \approx 0.1$ would give $\Delta\beta =
0.2$. The amount of extinction would be sufficient to provide an
intrinsic spectrum compatible with the Wind model. The extinction law
prescribed for the Small Magellanic Cloud bar is steeper than the LMC
case, and so if applicable should require less dust.

There cannot be a very large amount of extinction at the burst
redshift. To get the observed spectrum with as little extinction as
$A(V)=0.5$ requires a flat or rising intrinsic spectrum. Such an
intrinsic spectrum is inconsistent with the fireball model.

\citet{hpozk+05} present the {\em Swift}/XRT's X-ray upper limits,
from observations later than those presented here. We compare the
X-ray limits to the optical by extrapolating the $R$-band flux density
to their epoch, using our fitted models. We convert their flux limit
to a spectral flux density limit using the spectral information given,
both for their earliest epoch at $\sim 50$~ksec, and for the overall
observation. We find that the early optical spectral index of
$\beta=-0.8$ is marginally consistent with the X-ray limit,
overestimating the X-ray 90\% upper limit by approximately a factor of
3. (Given that the uncertainty is $\beta$ is 0.1, note that
$\beta=-0.9$ would only overestimate the X-ray limit by a factor of
1.5, and $\beta=-0.95$ would match it.) The expected initial intrinsic
$\beta=-0.8$ (ISM model; Table \ref{tab:alphabeta}) would be
marginally consistent, but the expected $\beta=-0.6$ (Wind; Table
\ref{tab:alphabeta}) would be inconsistent, overestimating the X-ray
limit by at least a factor of 14.

This X-ray--optical spectrum is consistent with the passage of
$\nu_c$. With an ISM model, $\nu_c$ has passed below the optical,
steepening the spectrum between $R$ and the X-ray at later times. With
a Wind profile, $\nu_c$ increases with time, placing it between $R$
and the X-ray at the time of the X-ray limit. However, $\nu_c$ would
not get far above the optical and the spectrum steepens between
$\nu_c$ and the X-ray. In both cases, the steepening $\Delta\beta =
-0.5$ above $\nu_c$ results in a model-predicted X-ray flux well below
the {\em Swift}/XRT limit.

\subsection{Fireball Model Constraints}\label{sec:fireball}

If the sub-luminosity of the earliest $R$-band observation (44~s) were
known to be due to the passage of the peak $\nu_m$, there would be
three numerical constraints for the fireball model's spectral
parameters: the peak time in $R$-band, the peak flux density value at
that time, and the cooling break time at the $R$-band frequency.  The
early lightcurve behavior is neither strong evidence for a rollover
from an earlier flatter lightcurve evolution, nor for a particular
physical mechanism for a rollover. Therefore the first two constraints
are inequalities; the peak $\nu_m$ could have have passed the optical
band earlier.

The observed decay from $44$ to $6000$~s requires that the passage
$\nu_m = \nu(R)$ is occurring or has occurred at $t \leq 44$~s. If
$\nu_m$ passed the $R$-band before $44$~s, the synchrotron spectral
peak would be at a lower frequency and a higher flux density than the
$R$-band's. Thus the peak flux density $f(\nu_m)$ at $44$~s must be
at least as bright as the initial $R$-band point. 

The peak passage $\nu_m$ is quite early, having occurred at a time no
later than 2.1 times the GRB duration measured by \citet{gm05}. The
cooling break passage time of $\approx 5700$~s is significant, placing
$\nu_c$ in the optical at a fairly early time. This is quite early in
comparison to several other cases where broadband data has indicated
$\nu_c$ is above the optical at a time $\sim$ days post-burst, e.g.,
970508 \citep{gwbgs+98}, 000301C \citep{bdfkb+01}, or 011211
\citep{jhfgp+03}. The implication in the ISM case with $\nu_c \propto
t^{-1/2}$ would be a low value of $\nu_c$ during the first
observation, with $\nu_c \sim 5\times10^{15}$Hz at $44$~s.

We consider the observational constraints in light of both the ISM and
the Wind models for the data. The synchrotron spectrum of a spherical
fireball expanding into the ISM at a known redshift depends upon five
parameters: energy $E$, density $n$, the power law slope of the
electron energy distribution $p$, the energy fraction partitioned to
the electrons, $\epsilon_e$, and the energy partition to the magnetic
fields, $\epsilon_B$ \citep[see][and references
therein]{piran04}. With three constraints and the value of $p$, we can
put three of $E$, $n$, $\epsilon_B$, and $\epsilon_e$ in terms of the
fourth, thus placing a limit upon a combination of two parameters.

We use the equations of \citet{gs02} for the ISM slow-cooling case for
$\nu_m$, $\nu_c$, and $F_{\nu_m}$. Adopting $p=2.55$, we find the
numerical relations using the redshift $z=3.793$ \citep{pefbc05} and a
cosmology with $\Omega_m=0.3$, $\Omega_{\Lambda}=0.7$,
$H_o=65$~km~s$^{-1}$Mpc$^{-1}$ to determine the appropriate luminosity
distance. We can then isolate
\begin{equation}
(\epsilon_e/0.1) (\epsilon_B/0.01)^{1/3} \leq 0.024,
\end{equation}
where $\epsilon_e$ and $\epsilon_B$ are scaled to 10\% and 1\%
respectively as these are reasonable expectations for the parameters
from broadband fits in many afterglow cases \citep[see,
e.g.,][]{pk02,yhsf03}. Within the simple fireball model that
well-describes this afterglow case in the optical-IR, the implication
is that the microphysical energy partitions must be quite small. For
example, if $\epsilon_e = 1\%$, then $\epsilon_B = 0.01\%$. Other
parameters can be put in terms of these, thus density $n \propto
\epsilon_B^{-10/6}$.  Keeping $n < 100$~cm$^{-3}$ requires
$\epsilon_B$ to be not much less than 1\%.

We note that the microphysical parameter combination in eq. 3 is
proportional to the peak flux density at 44~s as $F_{\nu_m}^{-1/3}$,
and to the frequency of the peak $\nu_m$ at 44~s as $\nu^{1/2}$. Thus
if the peak was significantly below the $R$ band at that time, this
combination of microphysical parameters would be notably smaller.

For a Wind model, the physical parameter set is the same except that
the density is parameterized by $A_{*}$, where $\rho =
5\times10^{11}A_{*}r^{-2}$~g~cm$^{-1}$. We do a similar analysis using
\citet{gs02}'s equations. In this case, we find 
\begin{equation}
(\epsilon_e/0.1) (\epsilon_B/0.01)^{1/3} \leq 0.08.
\end{equation}
Once again the microphysical parameters must be small (and increasingly small as
the initial value of $\nu_m$ is placed further below the $R$-band
frequency). 

There is an additional constraint:
\begin{equation}
A_{*} \geq 0.0004 (\epsilon_e/0.1)^{2}.
\end{equation}
This inequality indicates that, if $\nu_m < \nu(R)$ at the initial
observation, $A_{*} (\epsilon_e/0.1)^{-2}$ would be larger than
0.0004. However, from the previous relation the low $\nu_m$ would
require smaller microphysical parameters, $\epsilon_e$ or $\epsilon_B$
or both. If $\epsilon_e$ is decreased in the case $\nu_m < \nu(R)$,
then $A_{*}$ would remain low. A value $A_{*} = 0.0004$ is a very
small wind outflow parameter; $A_{*} \sim 1$ is observed in Wolf-Rayet
stars \citep[as reviewed by][]{willis91}. The physical parameter
constraints are somewhat easier to fulfill with the ISM fireball model
than with the Wind model.

\section{Conclusions}

We observe a shallow break of $\Delta\alpha \approx 0.3$ in the $R$
lightcurve of the GRB~050502A afterglow at approximately 6000~s
post-burst. We note that $J$, $H$, and $K_s$ observations during this
period, not as well-sampled, do not constrain the presence or absence
of such a lightcurve break in the NIR. With the shallow
$\Delta\alpha$, we conclude that the optical break represents the
passage of the synchrotron cooling frequency $\nu_c$ through $R$.

The observed spectral index is $\beta = -0.8 \pm 0.1$. This is
consistent with an ISM model for the broken $R$-band lightcurve, and
slightly steeper than expected in a Wind model. The temporal and
spectral index closure relations slightly favor an ISM over a Wind
interpretation. A small amount of dust at the host redshift would
steepen an intrinsically flatter spectrum sufficiently to accomodate
the Wind interpretation.

There is no evidence for an overluminosity at our earliest
observation, 44~s after the initial gamma-rays. The first
lightcurve point appears suppressed, but from a single point we cannot
conclude that this is the case. At a minimum, the early decay requires
$\nu_m$ at or below $\nu(R)$ at that time. Using that constraint and
the observation of $\nu_c$'s passage indicates that in both the ISM
and the Wind explanations small microphysical energy partitions are
required. The Wind interpretation expects an exceptionally low wind
outflow parameter $A_{*}$, which may again somewhat favor an ISM
interpretation.

\citet{gmgms+05} find multicolor evidence for a lightcurve bump at $t
\approx 1700-3400$~s. A single ROTSE data point during this period has
a slight rise above the overall decay, but not at a significant level
given its signal to noise. $JHK_s$ sampling is insufficient to address
whether the bump occurs at these wavelengths. Without the well-sampled
MDM $R$-band observations presented here from 5800-28000~s,
\citet{gmgms+05} did not detect the passage of $\nu_c$ through $R$.

\citet{gmgms+05} find the data favors a density variation over a
refreshed shock as the source of the bump. In that case, the
significance of the bump is greater than expected for $\nu(R) > \nu_c$
\citep[see][]{np03}. For the Wind case the cooling break passage
observed in the dataset presented here would be from $\nu(R) > \nu_c$
to $\nu(R) < \nu_c$ at $\approx$6000~s, so this again favors an ISM
over a Wind interpretation.

We note that the observed steeper, steady decay following the
lightcurve break is evident in the data over at least 0.7 logarithmic
decades in time. Despite the evidence for a density variation in the
data of \citet{gmgms+05}, such an impulsive event cannot explain the
sustained change in lightcurve evolution seen here.

There are several lines of evidence, including the spectral index, the
closure relations of spectral and temporal indices, and the optical
bump seen by \citet{gmgms+05}, that all somewhat favor an ISM model
over a Wind one for this afterglow.

With {\em Swift} and INTEGRAL providing rapid triggers, and rapid
response instruments such as ROTSE and PAIRITEL providing followup,
the GRB community is accumulating bursts with prompt, detailed
observations. Such data sets should further probe the physics of GRB
afterglows.

\acknowledgements

ROTSE-III has been supported by NASA grant NNG-04WC41G, NSF grant
AST-0407061, the Australian Research Council, the University of New
South Wales, the University of Texas, and the University of Michigan.
Work performed at LANL is supported through internal LDRD funding.
Special thanks to the observatory staff at McDonald Observatory,
especially David Doss.  

The MDM work has been supported by the
National Science Foundation under grant 0206051 to JPH. 

The Peters Automated Infrared Imaging Telescope (PAIRITEL) is operated
by the Smithsonian Astrophysical Observatory (SAO) and was made
possible by a grant from the Harvard University Milton Fund, the
camera loan from the University of Virginia, and the continued support
of the SAO and UC Berkeley.  Partial support is also supplied by a
NASA {\em Swift} Cycle 1 Guest Investigator grant.

The Digitized Sky
Surveys were
produced at the Space Telescope Science Institute under
U.S. Government grant NAG W-2166. The images of these surveys are
based on photographic data obtained using the Oschin Schmidt Telescope
on Palomar Mountain and the UK Schmidt Telescope. The plates were
processed into the present compressed digital form with the permission
of these institutions.

\newcommand{\noopsort}[1]{} \newcommand{\printfirst}[2]{#1}
  \newcommand{\singleletter}[1]{#1} \newcommand{\switchargs}[2]{#2#1}

\begin{deluxetable}{lcrrc}
\tablewidth{0pt}
\tablecaption{Optical Photometry for GRB~050502A\label{tab:photometry}}
\tabletypesize{\scriptsize}
\tablehead{
  \colhead{Telescope} &
  \colhead{Filter} &
  \colhead{$t_{\mathrm{start}}$ (s)} &
  \colhead{$t_{\mathrm{end}}$ (s)} &
  \colhead{Magnitude}
}
\startdata
ROTSE-IIIb & None &         44.0 &         49.0 & $14.28\pm 0.11$\\
ROTSE-IIIb & None &         59.0 &         78.9 & $14.32\pm 0.09$\\
ROTSE-IIIb & None &         88.5 &        108.0 & $14.71\pm 0.16$\\
ROTSE-IIIb & None &        117.2 &        136.8 & $15.08\pm 0.23$\\
ROTSE-IIIb & None &        146.7 &        180.5 & $15.63\pm 0.23$\\
ROTSE-IIIb & None &        190.0 &        239.6 & $15.92\pm 0.24$\\
ROTSE-IIIb & None &        248.7 &        327.7 & $16.19\pm 0.33$\\
ROTSE-IIIb & None &        337.5 &        416.7 & $16.58\pm 0.22$\\
ROTSE-IIIb & None &        426.5 &        546.1 & $17.04\pm 0.20$\\
ROTSE-IIIb & None &        555.3 &        754.4 & $17.30\pm 0.21$\\
ROTSE-IIIb & None &        763.6 &        962.7 & $17.50\pm 0.28$\\
ROTSE-IIIb & None &        971.9 &       1309.9 & $17.75\pm 0.24$\\
ROTSE-IIIb & None &       1319.7 &       1658.2 & $18.50\pm 0.25$\\
ROTSE-IIIb & None &       1667.4 &       2214.4 & $18.66\pm 0.13$\\
ROTSE-IIIb & None &       2224.1 &       2908.7 & $18.57\pm 0.14$\\
ROTSE-IIIb & None &       2918.0 &       3881.3 & $19.09\pm 0.11$\\
ROTSE-IIIb & None &       3890.5 &       5075.0 & $19.52\pm 0.19$\\
ROTSE-IIIb & None &       5084.1 &       6741.0 & $19.84\pm 0.22$\\
ROTSE-IIIb & None &       6750.1 &       8823.3 & $20.14\pm 0.18$\\
ROTSE-IIIb & None &       8832.5 &      11670.7 & $20.90\pm 0.29$\\
ROTSE-IIIb & None &      11680.6 &      15415.2 & $20.95\pm 0.27$\\
ROTSE-IIIb & None &      15425.1 &      20274.6 & $21.34\pm 0.33$\\
ROTSE-IIIb & None &      20284.1 &      28421.6 & $<21.67$\\
MDM & $R$ &       5816.0 &       6416.0 & $19.61\pm 0.03$\\
MDM & $R$ &       6458.0 &       7058.0 & $19.79\pm 0.03$\\
MDM & $R$ &       7105.0 &       7705.0 & $20.03\pm 0.03$\\
MDM & $R$ &       7750.0 &       8350.0 & $20.17\pm 0.03$\\
MDM & $R$ &       8386.0 &       8986.0 & $20.32\pm 0.03$\\
MDM & $R$ &       9025.0 &       9625.0 & $20.37\pm 0.04$\\
MDM & $R$ &       9928.0 &      10828.0 & $20.50\pm 0.04$\\
MDM & $R$ &      10871.0 &      11771.0 & $20.65\pm 0.04$\\
MDM & $R$ &      11812.0 &      12712.0 & $20.77\pm 0.06$\\
MDM & $R$ &      12750.0 &      13650.0 & $20.79\pm 0.08$\\
MDM & $R$ &      13687.0 &      14587.0 & $20.91\pm 0.06$\\
MDM & $R$ &      14621.0 &      15461.0 & $21.03\pm 0.06$\\
MDM & $R$ &      15645.0 &      16545.0 & $21.17\pm 0.06$\\
MDM & $R$ &      16587.0 &      17487.0 & $21.23\pm 0.06$\\
MDM & $R$ &      17537.0 &      18737.0 & $21.38\pm 0.06$\\
MDM & $R$ &      18797.0 &      19997.0 & $21.49\pm 0.06$\\
MDM & $R$ &      20115.0 &      21915.0 & $21.54\pm 0.05$\\
MDM & $R$ &      22041.0 &      23841.0 & $21.77\pm 0.06$\\
MDM & $R$ &      24057.0 &      25857.0 & $21.88\pm 0.08$\\
MDM & $R$ &      26358.0 &      28158.0 & $22.08\pm 0.11$\\
MDM & $R$ &      28319.0 &      30119.0 & $22.07\pm 0.19$\\
PAIRITEL & $J$ & 2322.5 & 3112.4 & $17.38\pm0.10$\\
PAIRITEL & $J$ & 4942.0 & 6335.4 & $18.56\pm0.21$\\
PAIRITEL & $J$ & 6365.4 & 7583.0 & $18.32\pm0.17$\\
PAIRITEL & $J$ & 8158.9 & 9903.5 & $<19.25$\\
PAIRITEL & $J$ & 9970.9 & 11722.3 & $<19.07$\\
PAIRITEL & $J$ & 11790.4 & 13538.4 & $<19.21$\\
PAIRITEL & $J$ & 13568.0 & 15316.8 & $<19.36$\\
PAIRITEL & $J$ & 15382.5 & 17134.5 & $<19.40$\\
PAIRITEL & $J$ & 17201.3 & 17979.7 & $<19.18$\\
PAIRITEL & $J$ & 18951.5 & 20742.1 & $<19.40$\\
PAIRITEL & $H$ & 2322.5 & 3112.4 & $16.64\pm0.10$\\
PAIRITEL & $H$ & 4942.0 & 6335.4 & $17.70\pm0.23$\\
PAIRITEL & $H$ & 6365.4 & 7583.0 & $17.77\pm0.23$\\
PAIRITEL & $H$ & 8150.3 & 9903.5 & $<18.55$\\
PAIRITEL & $H$ & 9970.9 & 11722.3 & $<18.19$\\
PAIRITEL & $H$ & 11790.4 & 13538.4 & $<18.30$\\
PAIRITEL & $H$ & 13568.0 & 15316.8 & $<18.32$\\
PAIRITEL & $H$ & 15382.5 & 17134.5 & $<18.42$\\
PAIRITEL & $H$ & 17201.3 & 17979.7 & $<18.10$\\
PAIRITEL & $K_s$ & 2322.5 & 3112.4 & $15.85\pm0.11$\\
PAIRITEL & $K_s$ & 4942.0 & 6335.4 & $16.67\pm0.18$\\
PAIRITEL & $K_s$ & 6365.4 & 7583.0 & $16.84\pm0.22$\\
PAIRITEL & $K_s$ & 8150.3 & 9903.5 & $17.36\pm0.31$\\
PAIRITEL & $K_s$ & 9970.9 & 11722.3 & $<17.33$\\
PAIRITEL & $K_s$ & 11790.4 & 13538.4 & $<17.35$\\
PAIRITEL & $K_s$ & 13568.0 & 15316.8 & $<17.57$\\
PAIRITEL & $K_s$ & 15382.5 & 17134.5 & $<17.74$\\
PAIRITEL & $K_s$ & 17201.3 & 17979.7 & $<17.21$\\
PAIRITEL & $K_s$ & 18951.5 & 20742.1 & $<17.74$\\
\enddata
\tablecomments{All times are in seconds since the burst time, 02:13:57 UT (see \S~\ref{sec:observations})}
\end{deluxetable}

\begin{deluxetable}{lccc}
\tablewidth{0pt}
\tablecaption{Fit Parameters and $\chi^2$\label{tab:fitpars}}
\tabletypesize{\scriptsize}
\tablehead{
  \colhead{Parameter} &
  \colhead{description} &
  \colhead{fit value} &
  \colhead{units}
}
\startdata
\multicolumn{4}{c}{Fits to the Individual Bands} \\
\hline
$\alpha_{R1}$ & early $R$ power law index & $-1.131 \pm 0.023$ & - \\
$\alpha_{R2}$ & late $R$ power law index & $-1.437 \pm 0.022$ & - \\
$t_{b}$ & $R$ break time & $5690 \pm 750$ & sec \\
$f_{0,R}(t_{b})$ & $f_{\nu}(R)$ at $t_b$ & $46.7 \pm 8.7$ & $\mu$Jy \\
$\kappa$ & ROTSE color term &  $-0.250 \pm 0.073$  & mag \\
$\chi^2_R$ & $\chi^2$ of $R$ fit/DOF & 44.9/38 & - \\
\hline
$\alpha_J$ & $J$ power law index & $-1.33 \pm 0.18$ & - \\
$F_J$($t_0$) & $f_{\nu}(J)$ of 1st IR epoch  &  $182 \pm 16$ & $\mu$Jy \\
$\chi^2_J$ & $\chi^2$ of $J$ fit/DOF & 8.0/8 & - \\
\hline
$\alpha_H$ & $H$ power law index & $-1.36 \pm 0.21$ & - \\
$F_H$($t_0$) & $f_{\nu}(H)$ of 1st IR epoch  & $229 \pm 21$ & $\mu$Jy \\
$\chi^2_H$ & $\chi^2$ of $H$ fit/DOF & 3.6/7 & - \\
\hline
$\alpha_{K_s}$ & $K_s$ power law index & $-1.17 \pm 0.17$ & - \\
$F_{K_s}$($t_0$) & $f_{\nu}(K_s)$ of 1st IR epoch  & $316 \pm 32$  & $\mu$Jy \\
$\chi^2_{K_s}$ & $\chi^2$ of $K_s$ fit/DOF & 3.4/8 & - \\
\hline
\hline
\multicolumn{4}{c}{Multiband Fit: NIR decay index = $\alpha_{R1}$} \\
\hline
$\alpha_{R1}$ & early power law index & $-1.137 \pm 0.022 $ & - \\
$\alpha_{R2}$ & late power law index ($R$ only) & $-1.435 \pm 0.022$ & - \\
$t_{b}$ & $R$ break time & $5830 \pm 250$ & sec \\
$f_{0,R}(t_{b})$ & $f_{\nu}(R)$ at $t_b$ & $45.2 \pm 2.8$ & $\mu$Jy \\
$\beta$ & spectral index $\nu^{\beta}$ & $-0.785 \pm 0.052$ & - \\
$\kappa$ & ROTSE color term &  $-0.263 \pm 0.066$  & mag \\
$\chi^2$ & $\chi^2$ of multiband fit/DOF & 65.7/66 & - \\
\hline
\hline
\multicolumn{4}{c}{Multiband Fit: NIR decay index = $\alpha_{R2}$} \\
\hline
$\alpha_{R1}$ & early power law index ($R$ only) & $-1.131 \pm 0.024$ & - \\
$\alpha_{R2}$ & late power law index & $-1.425 \pm 0.022$ & - \\
$t_{b}$ & $R$ break time & $5600 \pm 1800$ & sec \\
$f_{0,R}(t_{b})$ & $f_{\nu}(R)$ at $t_b$ & $47 \pm 21$ & $\mu$Jy \\
$\beta$ & spectral index $\nu^{\beta}$ & $-0.641 \pm 0.055$ & - \\
$\kappa$ & ROTSE color term &  $-0.252 \pm 0.096$  & mag \\
$\chi^2$ & $\chi^2$ of multiband fit/DOF & 68.2/66 & - \\
\enddata

\end{deluxetable}

\begin{deluxetable}{cccccc}
\tablewidth{0pt}
\tablecaption{Closure Relations for ($R$-band) Decay Index $\alpha$, Spectral Index $\beta$\label{tab:alphabeta}}
\tabletypesize{\scriptsize}
\tablehead{
  \colhead{Model} &
  \colhead{Relation (= 0)}\tablenotemark{a} &
  \colhead{Result} &
  \colhead{$p(\alpha)$} &
  \colhead{$p(\beta)$} &
  \colhead{expected $\beta(p(\alpha))$}
}
\startdata
ISM, initially $\nu_m < \nu(R) < \nu_c$  & $\alpha_1 -1.5\beta $  & $0.07 \pm 0.15$ & $2.51 \pm 0.03$ & $2.6 \pm 0.2$ & $-0.76 \pm 0.02$\\
ISM, initially $\nu_m < \nu(R) < \nu_c$  & $\alpha_2 -1.5\beta + 0.25 $ & $0.01 \pm 0.15$ &  $2.58 \pm 0.03$ & \tablenotemark{b} & $-0.79 \pm 0.02$\\
\hline
Wind, initially $\nu_m < \nu_c < \nu(R)$  & $\alpha_1 -1.5\beta + 0.25 $ & $0.32 \pm 0.15$  & $2.17 \pm 0.03$ & \tablenotemark{c} & $-0.59 \pm 0.02$\\
Wind, initially $\nu_m < \nu_c < \nu(R)$  & $\alpha_2 -1.5\beta + 0.5 $ & $0.26 \pm 0.15$ & $2.25 \pm 0.03$ &  $2.6 \pm 0.2$ & $-0.63 \pm 0.02$\\
\enddata
\tablecomments{The values of $\alpha_{1}$ is $\alpha_{R1}$, and $\alpha_{2}$ is $\alpha_{R2}$ from the Table \ref{tab:fitpars} fit to the individual bands. $\beta$ is $-0.8 \pm 0.1$ as determined in \S\ref{sec:beta}. The multiband fits also satisfy the closure relations given here.}
\tablenotetext{a} {as compiled in the review by \citet{piran04}}
\tablenotetext{b}{the value of $p$ is determined from $\beta$ assuming prebreak the optical and NIR have a spectrum $\propto \nu^{-(p-1)/2}$, and is the same postbreak}
\tablenotetext{c}{the value of $p$ is determined from $\beta$ assuming postbreak the optical and NIR have a spectrum $\propto \nu^{-(p-1)/2}$, and is the same prebreak}
\end{deluxetable}

\end{document}